\documentclass[11pt]{llncs}

\usepackage{fullpage}
\usepackage{graphicx}
\usepackage{algorithm2e}
\usepackage{subfigure}
\usepackage{epsfig}

\newcommand{\noop}[1]{}

\begin{document}

\title{Statistical Analysis of Privacy and Anonymity Guarantees\\ in Randomized Security Protocol Implementations}

\author{Susmit Jha\\  jha@cs.berkeley.edu}
\institute{Department of Electrical Engineering and Computer Science\\UC Berkeley}


\maketitle

\abstract{
Security protocols often use randomization to achieve probabilistic non-determinism. This non-determinism, in turn, is used in obfuscating the dependence of observable values on secret data. Since the correctness of security protocols is very important, formal analysis of security protocols has been widely studied in literature. Randomized security protocols have also been analyzed using formal techniques such as process-calculi and probabilistic model checking. In this paper, we consider the problem of validating implementations of randomized protocols. Unlike previous approaches which treat the protocol as a white-box, our approach tries to verify an implementation provided as a black box. Our goal is to infer the secrecy guarantees provided by a security protocol through statistical techniques. We learn the probabilistic dependency of the observable outputs on secret inputs using Bayesian network. This is then used to approximate the leakage of secret. In order to evaluate the accuracy of our statistical approach, we compare our technique with the probabilistic model checking technique on two examples: \emph{crowds protocol} and \emph{dining crypotgrapher's protocol}.
}

\section{Introduction}

A number of randomized protocols have been proposed to ensure the secrecy of certain facts which must not be disclosed while some consequences of these facts have to be made observable. There could be several motivations for this secrecy such as individual privacy expectations or negotiations among mutually untrusted parties. For example, a voting machine would be expected to retain the anonymity of the voter while recording his vote and hence, the vote recording protocol would do some periodic random reordering of votes. Other more involved examples of randomized protocols are contract-signing protocol, privacy-preserving auction protocol and  crowds protocol for routing messages. These protocols achieve this hiding of secret information by using randomization to obfuscate the relation between the secret data and the observable data. The goal of these protocols is to make it difficult for an attacker to learn the secret from the observable data. 

Since the correctness of security protocols is very important, formal verification of security protocols has been widely studied in literature. Deterministic protocols can be modeled as labeled transition systems and formal techniques such as model checking can be used for state exploration to verify properties expressed in temporal logic. Randomized security protocols can be modeled as discrete time Markov chains (DTMCs) or Markov decision processes (MDPs) and probabilistic model checking techniques can be used to verify properties expressed in stochastic temporal logic which is temporal logic augmented with probabilities. 

A key issue with these randomized protocols is that their implementations can be imperfect. For example, in the  \emph{Dining Cryptographers protocol}~\cite{golle-2004}, the coins being used by the cryptographers might be biased which might reveal probabilistic information about which cryptographer paid. Another example is that in the \emph{Crowds Protocol}~\cite{reiter-98}, the crowd might be infested by \emph{moles} which provide their observation to the adversary. This might be used by an adversary to guess the sender of a message with greater accuracy. 

The goal of this project is to develop a statistical technique for analyzing the protocol implementations and quantifying the anonymity loss. While trying to validate the implementation of the security protocols, we make the following assumptions.

\begin{enumerate}

\item The implementation is not created in an hostile environment and any implementation error is only an unintended bug such as use of poor pseudo-random generators. If the implementation is hostile, it can contain bugs which can not be easily detected by random sampling. For example: a crowds protocol implementation which would fail to anonymize for a particular path would not necessarily be detected as erroneous by our technique.

\item The implementation might have other vulnerabilities which make it possible to compromise it. Our analysis is limited to secrecy guarantees provided by the implementation and not to whether it is vulnerable to attacks.

\item The quality of the source of randomization while testing is the \emph{same} as when the implementation is deployed. If the source of randomization deteriorates and becomes more deterministic, the secrecy guarantees checked during testing will no longer hold true. 
\end{enumerate}

We assume that the implementation of the protocol is provided to us as a black-box. The reason we consider the implementation as a black box is because unlike protocols which are public, implementations could be an executable binary or a hardware implementation or an IP core and we would be able to check the implementation's correctness only by observing its inputs and outputs. 

Our approach to learn probabilistic dependencies is inspired from the work in reverse engineering genetic networks~\cite{pridgon-CIBCB04,liang-PSB98,tegner-NAS03} where a similar problem exists. Molecular and cellular processes form complex stochastic feedback networks where the regulatory molecules that control expression of genes are themselves controlled by other genes. Hence, an important problem is to reconstruct functional network architectures from the observed time series of gene expression data. This requires discovering probabilistic dependencies among the different genes. In verifying an implementation of a randomized protocol, we can obtain a number of sample traces (over both secret and observed values) of its working, and then use that to construct dependency of observed values from secret values. 

Once the probabilistic dependency of the observable values on the secret data has been learnt statistically from the traces of the randomized protocol implementation, the task of measuring the information leaked by the security protocol is similar to the channel capacity estimation problem. We compare the estimates obtained by our technique with that from probabilistic model checking to validate its accuracy. We use the two examples\footnote{models used for probabilistic model checking in \cite{kostas-JIC08} are available online at http://www.prismmodelchecker.org/}   used in \cite{kostas-JIC08} - dining cryptographers and crowds for comparison. 

The novel contributions made in this paper are -

\begin{enumerate}
\item This is a first completely statistical approach to verifying probabilistic security protocols and does not require any modeling of protocol as DTMC or MDP. This technique requires only sampling over the traces obtained by running the protocol implementation. We are not aware of any existing black-box testing approach for verifying security protocols.
\item We experimentally compare our technique with probabilistic model checking.
\end{enumerate}

The rest of the paper is organized as follows. In Section~\ref{rel}, we discuss related works. We present our statistical approach in Section~\ref{appr}. The experimental analysis of our approach is presented in Section~\ref{exp}. We mention some limitations of our work in Section~\ref{lim} and conclude by identifying some future work in Section~\ref{conc}.

\section{Related Works} \label{rel}
Characterization of secrecy loss in cryptographic protocols as channel capacity and mutual information has also been well studied in literature. Several efforts have also been made towards validation of randomized security protocols. In this section, we briefly summarize some of the relevant work.  

Secrecy provided by protocols is measured in terms of the entropy of the observable data in \cite{Serj-PET02}. This work uses the lack of information with the attacker as the notion of anonymity or secrecy. Thus, it implicitly assumes uniform distribution of the secret data. 

An alternative approach~\cite{moskowitz-WPES03} is to consider the mutual information between the observed data and the secret data. This is referred as the \emph{channel capacity} where the secrecy leak is modeled as the covert channel due to the imperfect nature of the security protocol. Consequently, this measure depends only on the protocol and not on the distribution of the secrets. Maximizing the channel capacity with respect to input distribution can be used as a measure of worst case secrecy loss. 

Different notions of quantitative information flow~\cite{diaz-PET02,edman-SY07} have been developed in context of analyzing information flow in a program. Many of these techniques assume some distribution (mostly uniform) on the secret input space and only provide guarantees for such input spaces. Clarkson, Myers and Schnieder~\cite{1068864} define information leaked as the difference between a prior guess of the attacker and the posterior belief of the attacker after the observation has been made. For example, an attacker could have some probabilistic model for the relation between secret $F$ and observable data $G$ as well as some guess about the secret facts $F$. After observing $G$, he can revise his guess. The incremental gain in knowledge of the attacker is defined to be the information flow from secret to observable data. The metric used to measure the difference in probability distribution in this work is the Kullback-Leibler distance.

Another approach based on hypothesis testing is described by Di Pierro et al~\cite{pierro-02}. Here attacks are seen to be like experiments conducted by an attacker to validate his hypothesis about the secret or to revise them. A similar approach by Charzikokolakis et al~\cite{chatzi-PP07} considers the problem of calculating errors of randomized protocols as finding how well can an attacker estimate the \emph{maximum a posteriori} rule given that a priori distribution is not known. They characterize this as Bayes risk.

Two existing approaches for verifying randomizing protocols that are most similar to our work are probabilistic variant of pi-calculus~\cite{even-ACM85,chatzi-TCS07} and probabilistic model checking~\cite{chatzi-TGC06}. We use probabilistic model checking as comparison point in our experimental evaluation. Probabilistic model checking~\cite{hinton-06} is an extension of model checking~\cite{burch-93} which was initially developed as a formal technique for finding bugs in circuits.  It is used to 
model and validate systems which exhibit stochastic behaviour.  Generally, probabilistic systems are specified as DTMCs or MDPs and the conditional probability of an observable value given a secret data is computed as the probability of reaching some state in the formal model. 

A major problem with application of this technique to verify implementations of randomized protocol is that one needs to abstract the implementation into a formal model. Even if the implementation is available as a source code or hardware design, it is non-trivial to derive the formal model from it. Errors might be overlooked or new errors might be introduced in derivation of the formal model from the implementation. In case, the implementation is only available as a black box, this technique can not be used.

\section{Statistical Approach} \label{appr}

In this section, we present our statistical approach to analysis of randomized protocols. We start with some preliminaries on information theory which is used in the rest of the section. We also briefly summarize the existing work on measuring loss of secrecy as channel capacity or mutual information to formally define our problem. We then show how we can use traces of the protocol implementation to learn probabilistic dependency graph between the secrets and the observable data. This dependency graph is basically a Bayesian network. We describe how we can learn the structure and parameters of the network from the traces and then, show how to estimate the mutual information from the learnt parameters. 

\begin{figure}[htp]
\centering
\includegraphics[width=60mm]{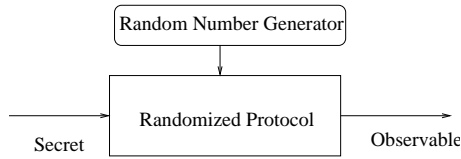}
\caption{Randomized Protocol Input/Output}\label{fig:randproto}
\end{figure}

In our approach, the randomized protocol to be analyzed can be viewed to be an information channel similar to Figure~\ref{fig:randproto}. It takes two inputs - the secret data $\mathcal{S}$ and a set of one or more random numbers $\mathcal{R}$. These are processed to output an observable event $\mathcal {O}$. The goal of the randomized protocol is to ensure that it is difficult to infer the value of secret data from the observable events. Thus, our task is to characterize \emph{how much} information about $\mathcal{S}$ is leaked through $\mathcal{O}$. 

\subsection{Information Theory}

Analysis of probabilistic systems relies on a number of tools and concepts used in information theory to reason about the uncertainty of some data and the amount of information it can reveal about some other data which was used in its computation. 

An important notion borrowed from information theory is that of \emph{entropy}. If $X$ is a random variable, then $H(X)$ denotes its entropy and is defined as $H(X) = - \sum_{x\in \mathcal{X}} p(x) log\;p(x)$ where $\mathcal{X}$ denotes the domain of $X$. Entropy measures the uncertainty of the random variable. If $log$ is taken with base 2, entropy is also the measure of number of bits required to express $X$. Intuitively, if $k$ bits are used to encode a variable, then it can take $2^k$ possible values assuming an uniform distribution. Thus, the entropy is measure of the information content in $X$.

Another important notion is that of \emph{relative entropy} between two probability distributions $p$ and $q$. It is defined as the Kullback-Leibler distance $D(p||q) = \sum_{x \in \mathcal{X}} p(x) log \; {p(x)}/{q(x)}$. This distance is always positive and is $0$ if $p=q$. Intuitively, it is measure of the expected difference in the number of bits required to code samples from $p$ when using a code based on $p$, and when using a code based on $q$. In coding theory, KL divergence is interpreted as the expected extra message-length per datum that must be communicated if a code that is optimal for a wrong distribution $q$ is used, compared to using a code based on the true distribution $p$.

A related concept is that of \emph{conditional entropy} of two random variables $X$ and $Y$. The conditional entropy $H(X|Y)$ is $H(X|Y) = - \sum_{y \in \mathcal Y} p(y) \sum_{x \in \mathcal X} p(x|y) log \; p(x|y)$ measures the uncertainty of $X$ after $Y$ is observed. It summarizes the extra information in $X$ which is not inferred from $Y$. It is maximum when $H(X|Y) = H(X)$, in which case the uncertainty of $X$ remains unchanged on observing $Y$. It is minimum at $0$ if there is deterministic function  $f(y) = x$.

The change in the uncertainty of $X$ on observing $Y$ is provided by the difference between entropy of $X$ and the conditional entropy of $X$ and $Y$. This quantity is called the \emph{mutual information} which is defined as $I(X;Y)= H(X)- H(X|Y)$. A little algebra shows that $I(X;Y) = I(Y;X)$. This is the main quantity of interest to us. 

\subsection{Channel Capacity}

Randomized protocols can be viewed as a \emph{lossy communication channel}~\cite{kostas-JIC08} from $S$ to $O$ and can be represented as a tuple $\langle \mathcal{S}, \mathcal{O}, p(\cdot|\cdot) \rangle$ where $p$ is the conditional probability distribution of observation given the secret. The mutual information between $\mathcal{S}$ and the observation $\mathcal{O}$ defines the flow of information across this channel. This definition of information flow does not enforce any probability distribution over the secrets $\mathcal{S}$. The maximum capacity is defined to be the maximum flow possible across this channel. Thus, channel capacity of the randomized protocol is $\max_{p(s)} I(\mathcal{S};\mathcal{O})$, that is,

			$$max_{p(s)}I(\mathcal S;\mathcal O) = max_{p(s)} \{ H(\mathcal S) - H(\mathcal S|\mathcal O) \} = max_{p(s)} \sum_{s} \sum_{o} [ p(s) p(o|s) log \;  { p(s|o) / p(s) } ]$$

where $s$ and $o$ are different values taken by the secret and observable variables. 

The above expression which is maximized over $p(s)$ is over two parameter distributions $p(o|s)$ and $p(s|o)$.

We note that algebraic manipulation would not decrease the degree of freedom of this expression and it would always be over two unknown parameters. For example: if we use $p(s|o) = p(o|s) p(s) / p(o)$ to rewrite the above expression as

			$$max_{p(s)}I(\mathcal S;\mathcal O) = max_{p(s)} \{ H(\mathcal S) - H(\mathcal S|\mathcal O) \} = max_{p(s)} \sum_{s} \sum_{o} [ p(s) p(o|s) log \;  { p(o|s) / p(o) } ]$$

The above expression is still over two parameters $p(s|o)$ and $p(o)$.

Our goal is to estimate the channel capacity of the randomized protocol which would characterize the maximum secrecy loss. For this we will need to estimate the above two distributions for a given randomized protocol. 

We choose to estimate $p(s|o)$ and $p(o|s)$ since 
\begin{itemize}
\item estimating $p(o)$ without the knowledge of $p(s)$ (except that we know that this distribution $p(s)$ would maximize the mutual information) would be difficult.
\item the above expression can be easily shown to be convex in $p(s|o)$ and $p(s)$ and hence maximizing the expression using these as parameters can be easily accomplished using gradient descent technique.
\end{itemize}
 For cases, where the distribution $p(s)$ which maximizes mutual information is known, the problem would be limited to only finding $p(o|s)$ since the joint probability distribution can be derived and the marginals and conditionals computed thereof. In general, it is not possible to analytically find $p(s)$ at which the maximum would be attained. 

\subsection{Learning $P(\mathcal O|\mathcal S)$}

Our approach towards learning the conditional distribution $p(o|s)$ is to construct a Bayesian network with nodes $\mathcal S \cup \mathcal O$ and edges from $\mathcal S$ to $\mathcal O$. A Bayesian network is a probabilistic graphical model that represents the probabilistic dependencies over a set of random variables. It is a graphical way to represent the factorization of the join probability distribution such that for each node, only a conditional probability table is maintained which gives the conditional probability of the random variable at that node given the values of the random variables at the parents' nodes. In this particular instance, for each observable variable, it depends on one or more secret variables which would be the parent of the observable variable. Efficient algorithms exist that perform inference and learning in Bayesian networks.

Since the protocol is only available to us as a black box implementation, we need to learn the edge connectivity of the corresponding Bayesian network to find what secrets determine what observable variables. This problem of learning structure of Bayesian networks from traces has been previously studied in literature in context of learning genetic networks~\cite{liang-PSB98,tegner-NAS03}. We adapt their approach to our problem and summarize it here. A more detailed discussion of this technique is available in ~\cite{liang-PSB98}. One domain knowledge that we exploit is that the secret variables and the observable variables do not have any intra-dependencies, that is, the observable variables are conditionally independent of each other given the secret variables. This reduces the search space of possible structures of the Bayesian network. 

The structure learning algorithm works by using the mutual information measures between the observable variables and the secret variables. The algorithm begins by identifying observable variables which depend on only one secret variable. If the mutual information $M(O_i,S_j)$ is same as $H(O_i)$, then clearly the observable variable $O_i$ depends only on $S_j$ and there is a single 
edge ending at $O_i$ - the one from $S_j$ to $O_i$. After identifying all nodes which have single edge, the algorithm proceeds to identifying observable variable nodes which depend on two secret nodes. The approach is similar - if the mutual information content $M(O_i,[S_j,S_k])$ is same as the entropy $H(O_i)$, then there are two edges ending at $O_i$ starting from $S_j$ and $S_k$. The process is continued till the maximum bound on the number of edges received as input. The algorithm is described below in Algorithm~\ref{alg:bn}. Conservatively max-degree bound can be specified to be the larger of the sizes of the secrets and observables. 

\vspace{0.5cm}

\begin{algorithm}[H]
\SetLine
\KwIn{Max-degree $d$, Secrets $\mathcal S$, Observables $\mathcal O$ and a set of traces $\mathcal T$ over these variables}
\KwOut{Mapping of observable to secrets}

$\mathcal O_r = \mathcal O$;

\ForEach{observable $O_j$} {
Calculate $H(O_j)$ from $\mathcal T$;
}
\ForEach{$k$ in $1$ to $d$}{
\ForEach{observable $O_j \in \mathcal O_r$}{
\ForEach{$k$-size candidate subset $\mathcal S_c$ of $\mathcal S$}{
Calculate $M(\mathcal S_c, O_j)$ from $\mathcal T$;
\If{ $M(\mathcal S_c, O_j) = H(O_j)$ }{
Remove $O_j$ from $\mathcal O_r$;
Record edge from all $S_i \in  \mathcal S_c$ to $O_j$;
}
}
}
Increment k;
}
assert($\mathcal O_r$ is empty);
\caption{Bayesian Network Structure Learning Algorithm}
\label{alg:bn}
\end{algorithm}

Once the Bayesian structure has been learnt, the conditional probability values for each observable variable is learnt using {\em maximum likelihood estimates} (MLE)~\cite{heckerman-tut95,daniel-icml04}. Let $\theta$ denote the unknown parameter of the conditional probability distribution, then the maximum likelihood estimate of $ \theta$ would be given by
$argmax_{ \theta} P(\mathcal T | \theta)$. An unbiased estimator for MLE is the frequency count. 

\subsection{Estimating $P(\mathcal S|\mathcal O)$ for mutual information}
The second parameter that we need to learn in order to compute the mutual information is $P(\mathcal S|\mathcal O)$. We present two approaches to compute this probability distribution. First, we consider a special case where the conditional probability distribution is symmetric and then, we show a more general approach to this problem. The special case was used in probabilistic model checking approach~\cite{kostas-JIC08}. The more general case is used in coding theory to infer correct codes from wrong codes.

\subsubsection{Exploiting row symmetry of $P(\mathcal O|\mathcal S)$ }
We consider the particular case in which the rows of $P(\mathcal O|\mathcal S)$ are permutations of each other. In this particular case,\\
$I(\mathcal S;\mathcal O) = H(\mathcal O) – H(\mathcal O|\mathcal S)$ where $H(\mathcal O|\mathcal S)  =   \sum_s p(s) H(\mathcal O|\mathcal S=s)$.\\
By row symmetry  $H(\mathcal O|\mathcal S=s) = H(\mathcal O|S=s') = H(\mathcal O_s)$ for all $s,s'$. \\
Thus, $I(\mathcal S; \mathcal O) = H(\mathcal O) - H(\mathcal O_s)$

So, in this special case, $P(\mathcal S| \mathcal O)$ need not be explicitly calculated and mutual information can be directly computed using the symmetry.

\subsubsection{Inference using gradient descent }
In general case, we need to use gradient descent techniques to compute the channel capacity by maximizing the mutual information over $p(\mathcal S)$.  An example of such a technique is Arimoto-Blahut (AB) algorithm. A detailed discussion of this algorithm and its extensions is presented in \cite{reza-ISIT04}. It is essentially a Bayesian approach where parameter is itself treated as a random variable. The unknown distribution parameters for the conditional probability  $p(\mathcal S| \mathcal O)$ and $p(\mathcal S)$ are treated as two different random variables (co-ordinates) over which the maximum needs to be attained. 

AB algorithm is an iterative technique with the following update rules -
\begin{itemize}
\item  $p^{i+1}(s|o) = p^i(s) p(o|s) / \sum_{s} p^{i}(s) p(o|s)$ 
\item  $p^{i+1}(s) =  ( \prod_{o} p^{i+1}(s| o)^{p(o|s)}  )  /  [ \sum_{s'} ( \prod_o p^{i+1}(s'|o)^ {p{(o|s')} } ) ]$
\end{itemize}

Since the mutual information expression is convex in both the parameters, the above iterative algorithm would eventually terminate with the correct answer that would correspond to maximum over possible $p(\mathcal S)$ and hence, is the channel capacity of the randomized protocol. 

\section{Experiments and Results} \label{exp}

We validated our statistical technique by comparing it with probabilistic model checking on two examples - dining crypographer's protocol and crowds protocol. These particular examples were chosen because their PRISM~\cite{hinton-06} models (MDPs) are available from the PRISM tool website~\footnote{http://www.prismmodelchecker.org/}.

\subsection{Dining Cryptographer}
Some $k$ number of cryptographers are dining . Either one of them or the host will pay the bill. They have agreed not to find out who pays the bill in case the bill was paid by one of the cryptographers. They only want to find out whether the host paid the bill or one of the cryptographers did.  They each toss a coin and if their coin matches with the one on their left, they say 1 if they are paying and 0 if they are not. If the coins don't match they say 0 if they are paying and 1 if they are not. If no cryptographer paid, final exclusive or of their announcements would be $(a_1 \oplus a_2) \oplus \ldots (a_i \oplus a_{i+1}) \ldots (a_k \oplus a_1)$ zero. If a cryptographer paid: final exclusive or will yield 1. Thus, this protocol provides a channel to allow announcement from a sender while maintaining his anonymity. But this secrecy guarantee relies on the fairness of the coin used by the cryptographers. An attacker can bias the coins used by cryptographers and then try to learn which cryptographer paid.

\begin{figure}
     \centering
     \subfigure[k=3]{
     \includegraphics[width=.45\textwidth]{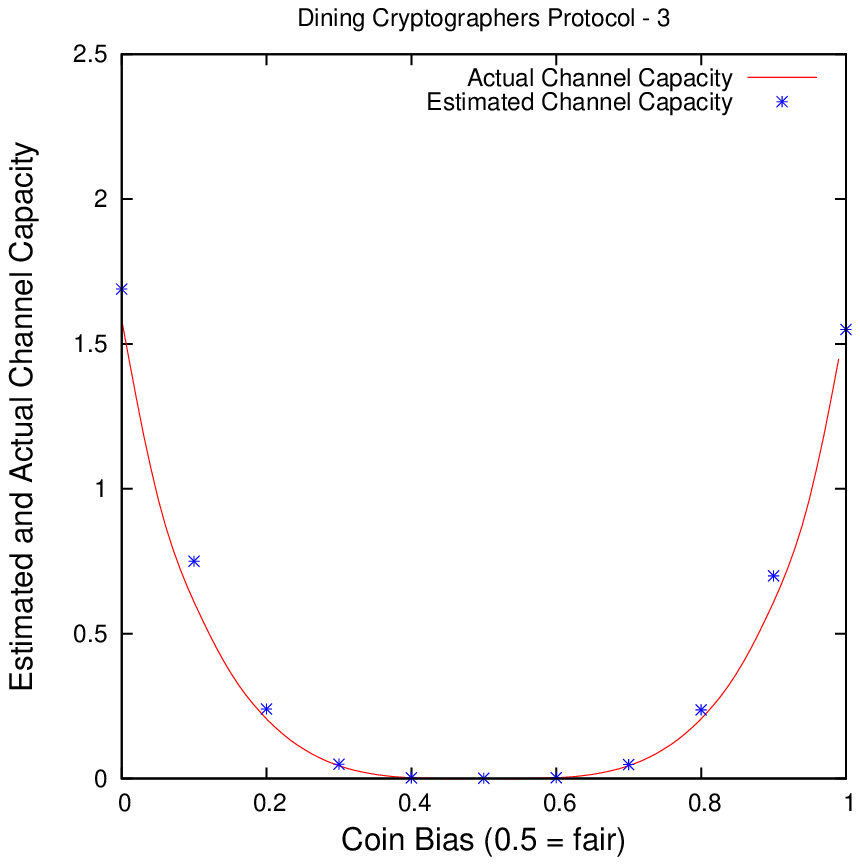}}
\hspace{0.1in}
     \subfigure[k=4]{
     \includegraphics[width=.45\textwidth]{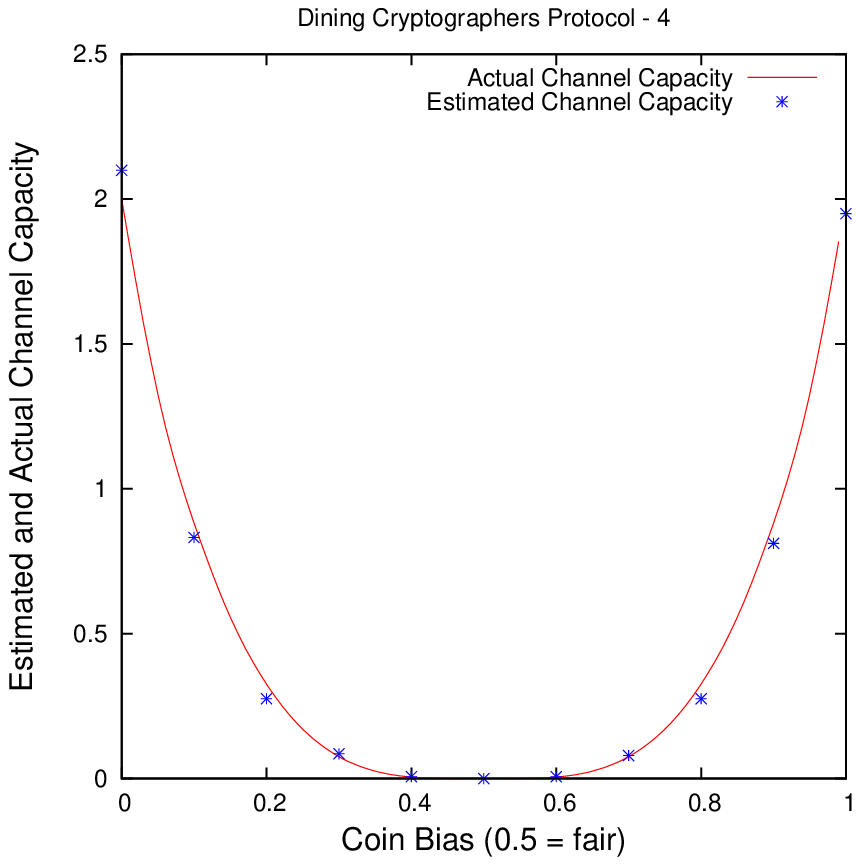}}
 \vspace{0.1in}
     \subfigure[k=5]{
     \includegraphics[width=.50\textwidth]{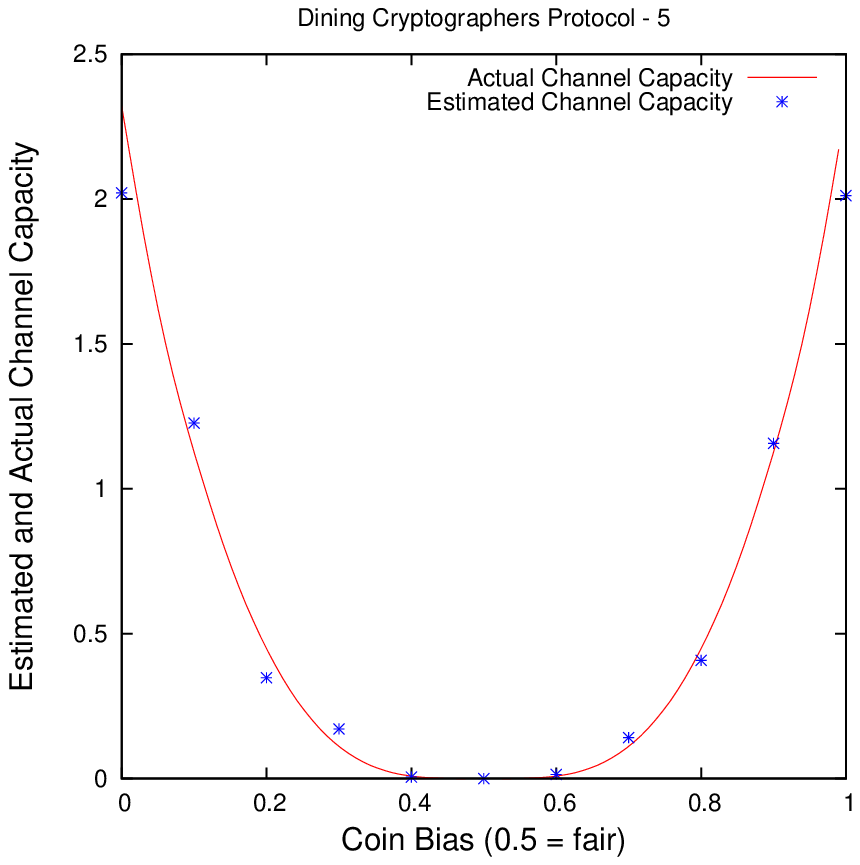}}
     \caption{Dining Cryptographers with varying number of cryptographers (k)}
     \label{fig:dc}
\end{figure}

We conducted 3 different set of experiments with $k=3,4,5$ for the above protocol. In each set of experiment, we considered the coins with heads probability varying from $0.0$ to $1.0$ in increments of $0.1$. We estimated the channel capacity for the protocol using probabilistic model checking and our technique. The corresponding plots are presented in Figure~\ref{fig:dc}. The plot shows the channel capacity estimated using probabilistic model checking as a continuous line while the statistically inferred data is shown as points. The plot shows that for this case study, the statistical estimates are very close to real values of channel capacity.

The channel capacity is observed to be minimum, that is 0, when fair coin is used by the cryptographers. This corresponds to the strong anonymity guarantees of the protocol if it is implemented using a fair coin. As the coin is made unfair, channel capacity rises and attains its maximum value which is close to log of the number of cryptographers when the coin is completely biased.

\subsection{Crowds Protocol}

\begin{figure}
     \centering
     \subfigure[1 Corrupt node]{
     \includegraphics[width=.45\textwidth]{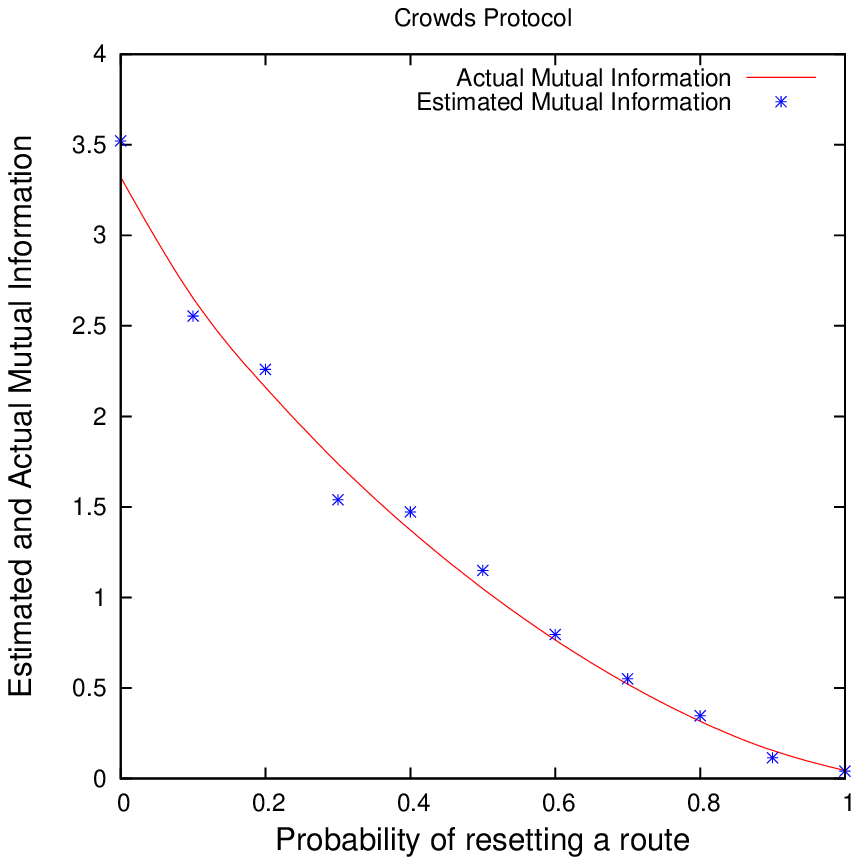}}
\hspace{0.1in}
     \subfigure[2 Corrupt nodes]{
     \includegraphics[width=.45\textwidth]{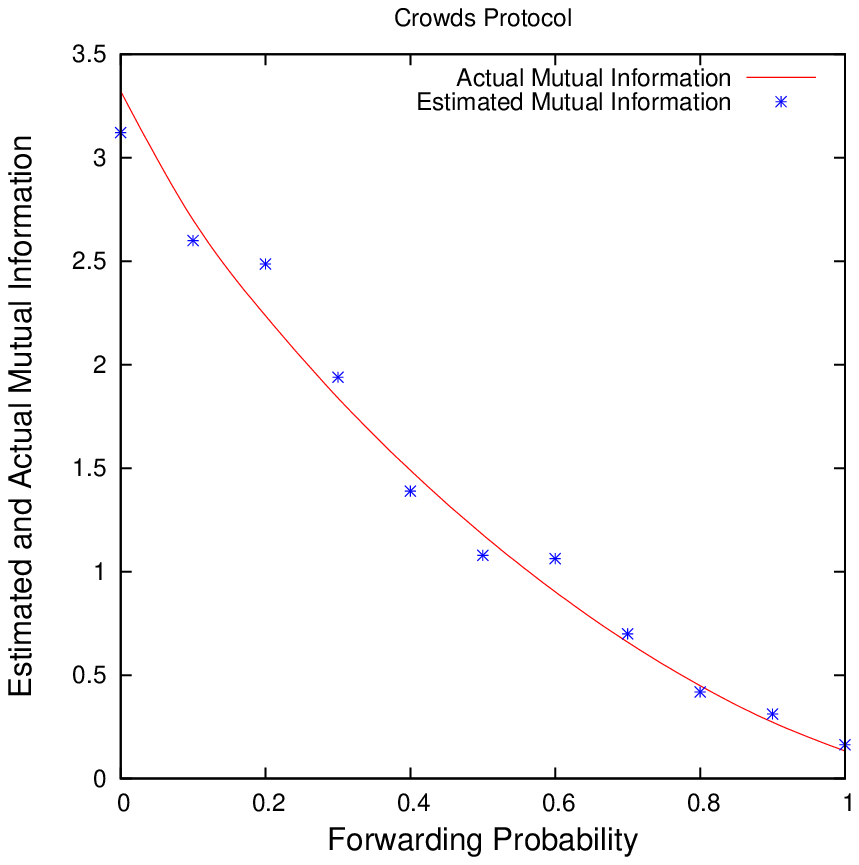}}
 \vspace{0.1in}
     \subfigure[5 Corrupt nodes]{
     \includegraphics[width=.45\textwidth]{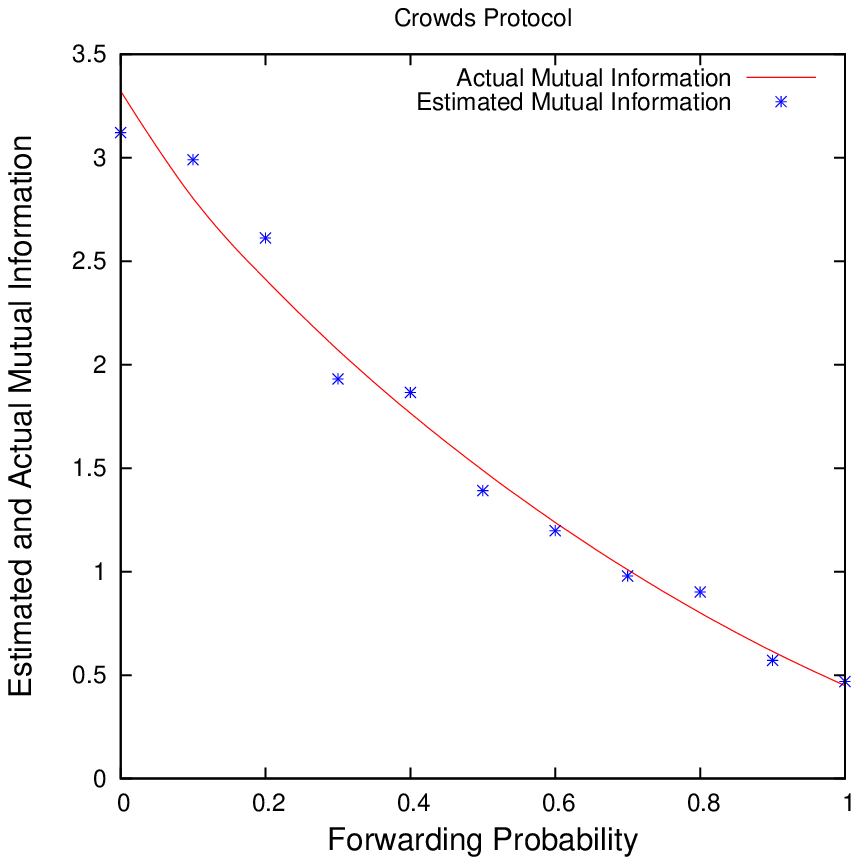}}
\hspace{0.1in}
     \subfigure[10 Corrupt nodes]{
     \includegraphics[width=.45\textwidth]{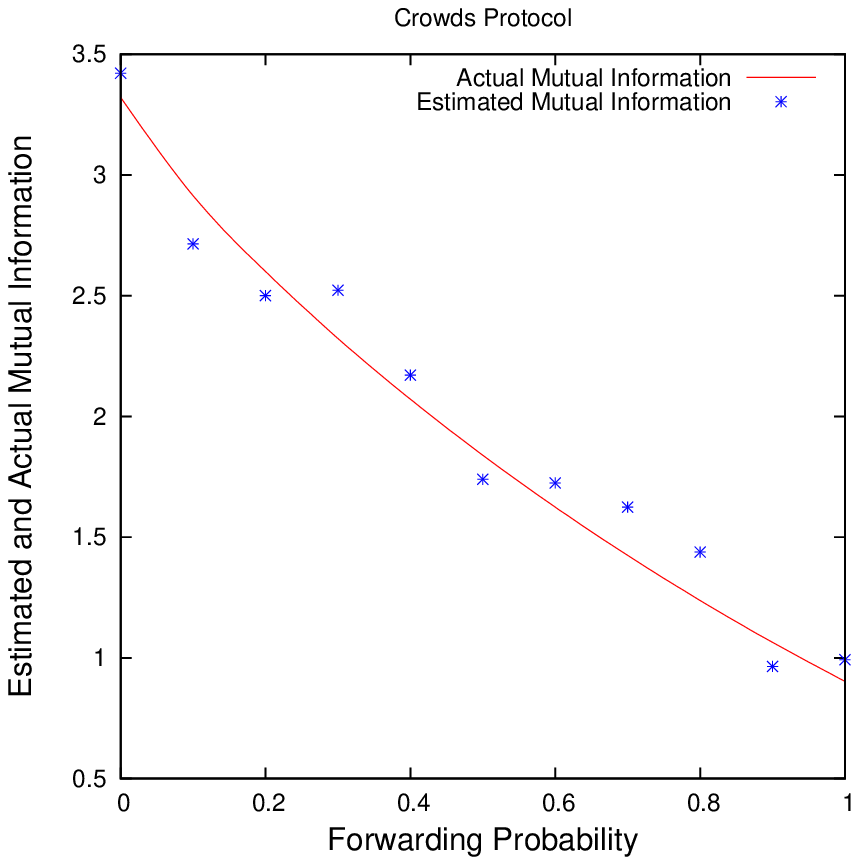}}
     \caption{Crowds Protocol with 10 honest nodes}
     \label{fig:crowd1}
\end{figure}

\begin{figure}
     \centering
     \subfigure[10 Corrupt node]{
     \includegraphics[width=.45\textwidth]{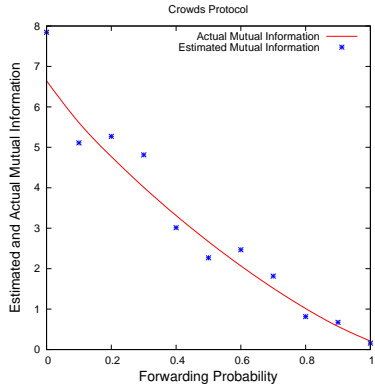}}
\hspace{0.1in}
     \subfigure[20 Corrupt nodes]{
     \includegraphics[width=.45\textwidth]{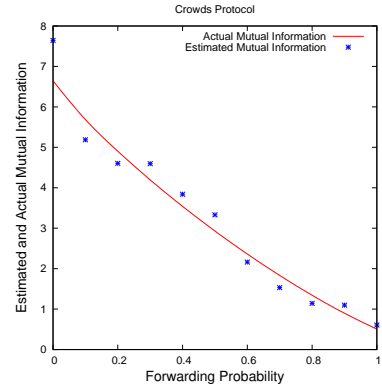}}
 \vspace{0.1in}
     \subfigure[50 Corrupt nodes]{
     \includegraphics[width=.45\textwidth]{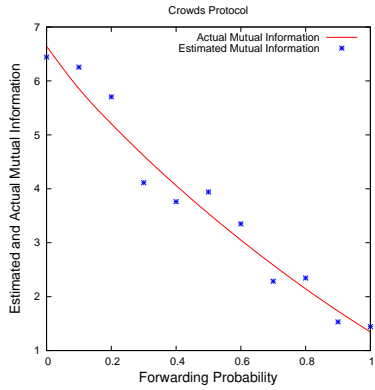}}
\hspace{0.1in}
     \subfigure[100 Corrupt nodes]{
     \includegraphics[width=.45\textwidth]{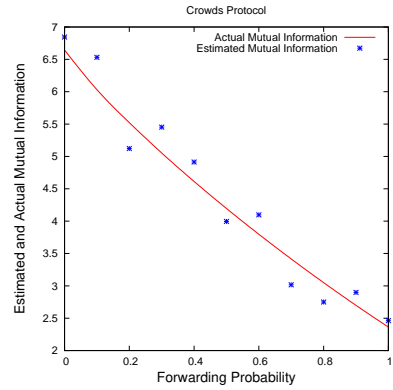}}
     \caption{Crowds Protocol with 100 honest nodes}
     \label{fig:crowd2}
\end{figure}

Crowds: This is meant to ensure that a message is transmitted from source to destination without revealing the identity of the source.  Initiator randomly selects a node to send it message. Each forwarding node, with some fixed probability delivers message to destination or sends to some other randomly selected node. Let some nodes be dishonest. They do not forward their messages and record the node which came before them in the path. As paths are rebuilt, the corrupt nodes will very likely see the sender node as predecessor more often than other nodes.  Plots in Figure~\ref{fig:crowd1} and Figure~\ref{fig:crowd2} assume that the attacker is on the path  from original source to destination.

We consider two different sets of experiments - the first set with 10 honest nodes and varying number of dishonest nodes (1,2,5,10) and the second set with 100 honest nodes and varying number of dishonest nodes (10,20,50 and 100). We plot the forwarding probability with the mutual information. While our statistical estimates are close to the values computed by probabilistic model checking for the smaller crowds, the variation increases for larger crowds illustrating that sampling was not uniform. 

The plots show that the maximum information is obtained when the forwarding probability is minimum. The maximum information correspond to base 2 logarithm of the total number of honest nodes in the crowd, that is, all the identities are revealed.

\section{Limitations} \label{lim}

We identify some key weaknesses of our approach.

\begin{enumerate}

\item We need to evaluate our technique on non-symmetric protocols. Currently AB algorithm is not implemented and hence, we can only work with symmetric protocols.

\item We assume that the source of randomization is of the same quality during testing and deployment. Most errors in implementations of security protocol arise due to poor randomization in deployment.

\item In cases, where we can identify that anonymity guarantees provided by an implementation is not sufficient, we do not provide any information regarding what was wrong with the implementation. In contrast, probabilistic model checking can localize errors in a protocol.

\item Unlike formal techniques which provide guarantees about their correctness, our statistical approach lacks any guarantee. It relies heavily on randomized sampling of the traces of the implementation.

\end{enumerate}

\section{Conclusion and Future Work} \label{conc}

The statistical approach to analyze security protocols seems promising. We plan to compare our technique more extensively with probabilistic model checking as well as other techniques to analyze randomized protocols. We identify some directions in which further work can be done on this project -
\begin{enumerate}
\item Implementations can be partially observable and hence, we can use more than just the secret and observable variables in our traces for more involved protocols.

\item This technique is not limited to security protocols and can also be extended to quantify the information flow in a program.

\item This analysis technique can be used to identify protocol parameters for some intended degree of secrecy. For example: how many nodes in a crowd should be corrupt such that the protocol is exactly k-anonymous.

\end{enumerate}

\section{Acknowledgment}
We are thankful to David Wagner for his comments on this work which started as a course project in his graduate course on Computer Security in Fall, 2008. I am thankful to Martin Wainwright for insightful suggestions about use of statistical techniques. I am also thankful to Sanjit Seshia for fruitful discussions. 

\bibliographystyle{plain}

\bibliography{proposal}

\end{document}